# Discriminative transition sequences of origami metamaterials for mechano-logic

Zuolin Liu[a,b], Hongbin Fang[a,1], Jian Xu[a], and K.W. Wang[b]

[a] Institute of AI and Robotics, Fudan University, Shanghai, 200092, China;
[b] Department of Mechanical Engineering, University of Michigan, Ann Arbor, MI 48109, USA;
[1] To whom correspondence may be addressed. Email: fanghongbin@fudan.edu.cn

**Abstract (250)**

Transitions of multistability in structures have been exploited for various functions and applications, such as spectral gap tuning, impact energy trapping, and wave steering. However, a fundamental and comprehensive understanding of the transitions, either quasi-static or dynamic transitions, has not yet been acquired, especially in terms of the sequence predictability and tailoring mechanisms. This research, utilizing the stacked Miura-ori-variant (SMOV) structure that has exceptional multistability and shape reconfigurability as a platform, uncovers the deep knowledge of quasi-static and dynamic transitions, and pioneers the corresponding versatile formation and tuning of mechanical logic gates. Through theoretical, numerical, and experimental means, discriminative and deterministic quasi-static transition sequences, including reversible and irreversible ones, are uncovered, where they constitute a transition map that is editable upon adjusting the design parameters. Via applying dynamic excitations and tailoring the excitation conditions, reversible transitions between all stable configurations become attainable, generating a fully-connected transition map. Benefiting from the nonlinearity of the quasi-static and dynamic transitions, basic and compound mechanical logic gates are achieved. The versatility of the scheme is demonstrated by employing a single SMOV structure to realize different complex logic operations without increasing structural complexity, showing its superior computing power and inspiring the avenue for efficient physical intelligence.

**Significance Statement (50-120)**

We pioneer an investigation of the quasi-static and dynamic transitions among multiple stable configurations, crucial for many functions, in a stacked Miura-ori-variant (SMOV) structural platform that has exceptional multistable and multidirectional reconfigurable features for functional adaptability. With innovation that reveals the predictability and discriminability of the transition sequences and the mechanisms for triggering different transitions, significant new knowledge of the transition mechanics is created, and novel strategies are fostered for tuning the multistable metamaterials for diverse functionalities. Moreover, the unique mechanical computing strength of the transition sequences for conducting logic operations is uncovered. Basic and compound logic gates are achievable with a single SMOV structure, which is a significant breakthrough of mechano-logic in terms of structural simplicity and functional versatility.

## Main Text

### Introduction

With the unique merit of exhibiting variable spectral gaps at different stable configurations, multistable mechanical metamaterials have facilitated extensive functions and applications, including phononic bandgap tuning (1–3) and broadband vibration control (4). Among these practices, the multistable metamaterials, which are fundamentally nonlinear in their constitutive profiles, are mainly operating in linear regimes within small deformations around different stable equilibria between configuration transitions. On the other hand, other prospects, such as nonreciprocal wave transmission (5–7), impact energy trapping (8), shock isolation (9), and transition signal propagation (10,11), have leveraged the nonlinear feature of global multistability, particularly the snap-through transitions among different stable configurations. Recently, there is a growing interest in harnessing multistability for mechanical logic gates (12–14) by correlating the mechanical configurations with their digital counterparts. Upon external inputs, the logic operation is determined by the sequence of configuration transitions. While these outcomes are intriguing, the current state-of-the-art technology mainly exploited transitions in an ad-hoc manner, and the underlying mechanics of a transition sequence and the corresponding triggering methods are often not well understood. In other words, systematic and comprehensive investigations into the global transition sequences have not been pursued, which is a major bottleneck that severely limits the robust realization of the many rich functions of multistability.

As a design motif, origami, the ancient art of transforming flat sheets into a sophisticated sculpture through folding, provides tremendous potentials in building multistable mechanical metamaterials owing to its large design space and intrinsic geometric nonlinearity (15). Foreseeable applications include mechanical memory devices (16), mechano-logic (14,17,18), and robotics (19). Recently, by incorporating multiple stacked Miura-ori units via a novel stacking strategy (21,22), a new "stacked Miura-ori-variant (SMOV)" structure is created. With exceptional multistability in inclined and curved directions and multiple configurations, the SMOV becomes a strong candidate for developing smart mechanical metamaterials with strong directional, configurational, and functional adaptability. Moreover, with 4 to 8 different stable configurations in a single SMOV cell, rich transition sequences are expectable, which brings fresh vitality to the creation of new functions, such as mechano-logic with strong computing capability.

With the abovementioned critical needs in advancing the knowledge of transitions in mechanical multistable metamaterials and the attractive features of SMOV, in this research, our goal is to utilize the SMOV structure as a podium for studying the rich multistability transition behaviors, understanding the underlying physics, and manipulating and harnessing the transition sequences. While this is an exciting opportunity, the complexity of the sequence also brings about major research challenges for us to address, so we can better exploit and leverage the underlying mechanisms of the transitions. Particularly, when transiting the SMOV from one stable configuration to another under rigid-folding, the kinematic bifurcation point (20) (aka, the kinematic singular state) will always be encountered due to the synchronous folding of the constituent cells as a single-degree-of-freedom mechanism. At this point, the SMOV has multiple transition paths via changing the folding direction, which exacerbates the difficulty in elucidating the transition sequence. As a consequence, the subsequent folding of the SMOV becomes indeterminant and unpredictable, which prevents the realization of various SMOV functionalities.

To achieve our research goal, we advance the state of the art by addressing the abovementioned challenges and pioneering an investigation of the quasi-static and dynamic transitions among the SMOV multistable configurations. First, we introduce flexibility into the connection between adjacent constituent units, which relaxes the strict rigid-folding kinematic constraints and allows each unit to deform asynchronously, thus making the transition sequence predictable by avoiding the kinematic bifurcation point. In addition, through systematic analysis of the quasi-static configuration switches, transition maps composed of reversible and irreversible transition sequences are revealed. Such transition maps can be further edited by engineering the design parameters of the SMOV structure. Configuration switches can also be triggered by dynamic excitations, in the form of steady-state oscillations around different stable states. Different from the quasi-static scenario, dynamic transitions between any of the two stable equilibria are always reversible, generating a bi-directional full-connected transition map.

Building on this foundation, we discover that the SMOV discriminative transition sequences, including quasi-static and dynamic maps, provide a novel platform for versatile logic operations. Rather than the conventional mechano-logic that a specific structure can only act as a single type of logic gate (12, 23, 24), the proposed multistable SMOV structure, as a novel element for logic operation, can serve as multiple types of logic gates. Moreover, instead of integrating multiple cells in conventional mechano-logic approaches, our scheme by incorporating a reservoir process can perform compound logic operations based on a single multistable SMOV cell, without increasing structural complexity. These findings, therefore, will inspire the avenue for mechanical intelligence to be harnessed in many systems, e.g., smart materials, MEMS, and robotics.

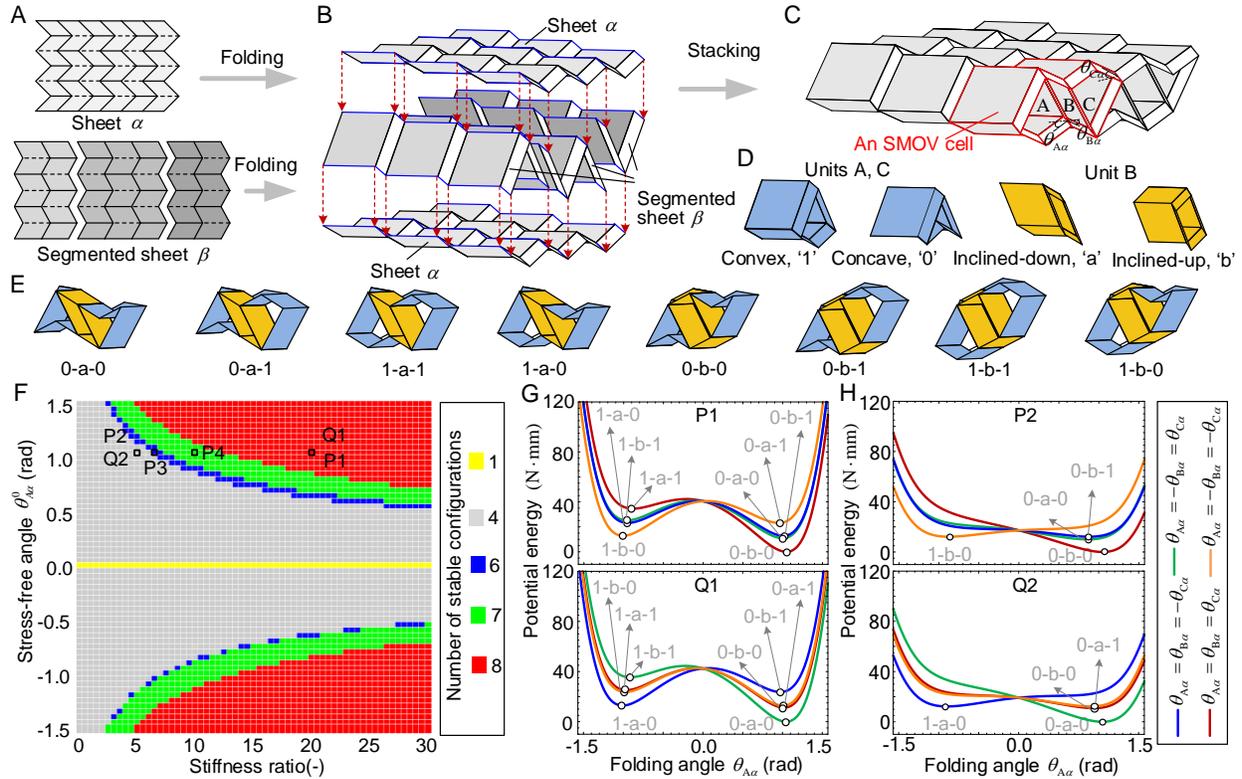

**Figure 1.** (*A*) Two Miura-ori sheets $\alpha$ and $\beta$ for constructing the multistable metamaterial. (*B*) Illustration of the stacking method. (*C*) A single layer of the SMOV metamaterial, in which a constituent cell, i.e., a SMOV cell, is highlighted. The SMOV cell is made up of three units, A, B, and C; their kinematics are governed by folding angles $\theta_{A\alpha}$, $\theta_{B\alpha}$, and $\theta_{C\alpha}$, respectively. (*D*) Different configurations of the units. (*E*) Eight different configurations of the SMOV cell. (*F*) Correlation between the number of stable configurations of a SMOV cell and the design parameters (stiffness ratio and stress-free angle). (*G*) and (*H*) Potential energy landscapes of the SMOV cell corresponding to points P1, Q1, P2, and Q2 in (*F*).

## Results

### The multistable Miura-variant metamaterial

The Miura-variant metamaterial utilized in this study is constructed by stacking two different Miura-ori sheets, $\alpha$ and $\beta$ (*Fig. 1A*) as presented in *Fig. 1B*, which includes a large number of tubular cells (*Fig. 1C*). Considering the periodicity, a basic constituent cell of the metamaterial, i.e., a stacked Miura-variant cell, is made up of three units, denoted by A, B, C and are highlighted in *Fig. 1C*; their folding motions can be uniquely described by the folding angles $\theta_{A\alpha}$, $\theta_{B\alpha}$, and $\theta_{C\alpha}$. Among them, units A and C are conventional stacked Miura-ori (SMO) units, which possess two different types of configurations, namely, the convex configuration ($\theta_{A\alpha} < 0$ and $\theta_{C\alpha} < 0$) and the concave configuration ($\theta_{A\alpha} > 0$ and $\theta_{C\alpha} > 0$); the newly generated unit, located between units A and C, can also achieve two different types of configurations, the inclined-up ($\theta_{B\alpha} > 0$) and the inclined-down ($\theta_{B\alpha} < 0$) configurations (*Fig. 1D*). Therefore, a single Miura-variant cell can exhibit eight different types of configurations by reconfiguring the constituent units (*Fig. 1E*). In what follows, for clarity, binary codes '1' and '0' are used to represent the convex and concave configuration of units A and C, respectively; 'a' and 'b' are adopted to denote the inclined-down and inclined-up configuration of unit B, respectively. Detailed kinematics of a single cell is presented in *SI Appendix, Section S1*.

The stability characteristics of a Miura-variant cell are determined by three design parameters: the stiffness ratio, defined as the ratio of the crease torsional spring stiffness per unit length of sheet $\alpha$ ($k_\alpha$) to that of sheet $\beta$ ($k_\beta$), the stress-free configuration of the cell when there is no internal force, and the corresponding stress-free folding angle (denoted as $\theta_{A\alpha}^0$, $\theta_{B\alpha}^0$, $\theta_{C\alpha}^0$). By tailoring these design parameters, the potential profile of a Miura-variant cell could exhibit different numbers of local minimum, corresponding to different numbers of stable configurations (see detailed derivations of the potential energy in *SI Appendix, Section S2*). For example, by setting the stress-free configuration at '0-b-0' and allowing the stiffness ratio and the stress-free angle $\theta_{A\alpha}^0$ to vary, the constituent cell could achieve 1, 4, 6, 7, or 8 stable configurations (*Fig. 1F*). For each point on the parameter plane, considering the binary configurations of units B and C, four potential energy curves can be plotted with respect to the folding angle of unit A (i.e., $\theta_{A\alpha}$). For instance, at point P1, all the four curves show prominent double-well characteristics, giving rise to eight stable configurations (*Fig. 1G, top*). By reducing the stiffness ratio, the potential wells with relatively shallow depths would disappear, thus reducing the number of stable configurations. Particularly, at point P2, all the four energy curves become mono-stable, producing four stable configurations (*Fig. 1H, top*); and at the line with zero stress-free angles (i.e., $\theta_{A\alpha}^0 = \theta_{B\alpha}^0 = \theta_{C\alpha}^0 = 0$), regardless of the stiffness ratio, the four curves completely coincide and share one potential well, which corresponds to the unique stress-free stable configuration. Examples of energy curves with 7, 6, and 1 stable configuration are given in *SI Appendix, Fig. S2C-E*, and evolution of the folding angles at the stable configurations with respect to the stiffness ratio and the stress-free angle is described in *SI Appendix, Fig. S2A*, and *Fig. S2B*, respectively.

Moreover, it is worth noting that even with the same number of stable states, the specific shapes of the stable configurations are still tunable by adjusting the stress-free configuration. For instance, with the same stiffness ratio and stress-free angle but different stress-free configurations ('0-b-0' at point P2 and '0-a-0' at point Q2), although the number of stable states remains four, the specific shapes of the stables configuration are not identical, changing from '1-b-0', '0-b-1', '0-a-0', '0-b-0' (*Fig. 1H, top*) to '1-a-0', '0-a-1', '0-b-0', '0-a-0' (*Fig. 1 H, bottom*). Similarly, by switching the stress-free configuration from '0-b-0' (point P1) to '0-a-0' (point Q1), the Miura-variant cell remains octa-stable, but the potential energy levels corresponding to the eight stable configurations are changed. Actually, for the Miura-variant cell, the number of stable states can be uniquely determined by the stiffness ratio and the stress-free angles, while the specific shapes of the stable configurations and the related potential energy levels also depend on the stress-free configuration. We will show later that in addition to modifying the overall potential profile of the Miura-variant cell, the three design parameters play a key role in governing the sequences of configuration transitions.

### Quasi-static transition sequences

Under the rigid-folding scenario, the kinematic constraints $\theta_{A\alpha} = \pm\theta_{B\alpha} = \pm\theta_{C\alpha}$ have to be precisely satisfied, which forces the three units of the SMOV cell to deform synchronously. Hence, a kinematic bifurcation point with $\theta_{A\alpha} = \theta_{B\alpha} = \theta_{C\alpha} = 0$ will always be encountered when transforming the cell among its stable configurations. When passing through this bifurcation point, the sign of the folding angle of each unit cannot be uniquely determined, which makes the transition sequences unpredictable. However, in practice, rigid-foldability cannot be perfectly satisfied due to the inevitable flexibility of the facets and creases, which relaxes the rigid-folding kinematic constraints by allowing each unit to deform asynchronously. Nevertheless, the folding of the adjacent units is not fully independent either; the connecting facets or creases will still impose certain constraints to restrict the folding differences between adjacent units. Specifically, to quantify such imperfect constraints between adjacent units A and B, as well as units B and C in the SMOV cell, two equivalent stiffness $k_1^*$ and $k_2^*$ are introduced; they are applied to the dihedral-angle differences between adjacent units. The newly introduced equivalent stiffness brings about additional potential energy (see detailed derivations in *SI Appendix, section S3*), which could fundamentally alter the overall potential energy landscape of the SMOV cell. Hence, starting from an initial configuration of the SMOV cell under displacement control, the path corresponding to the minimum energy can be searched via an optimization process. It is shown that with imperfect constraints, the kinematic bifurcation point is no longer encountered when transiting among the stable configurations, thus making the transition sequence deterministic and predictable. Actually, the transition sequence can be uniquely determined by locating the local minima on the energy landscape.

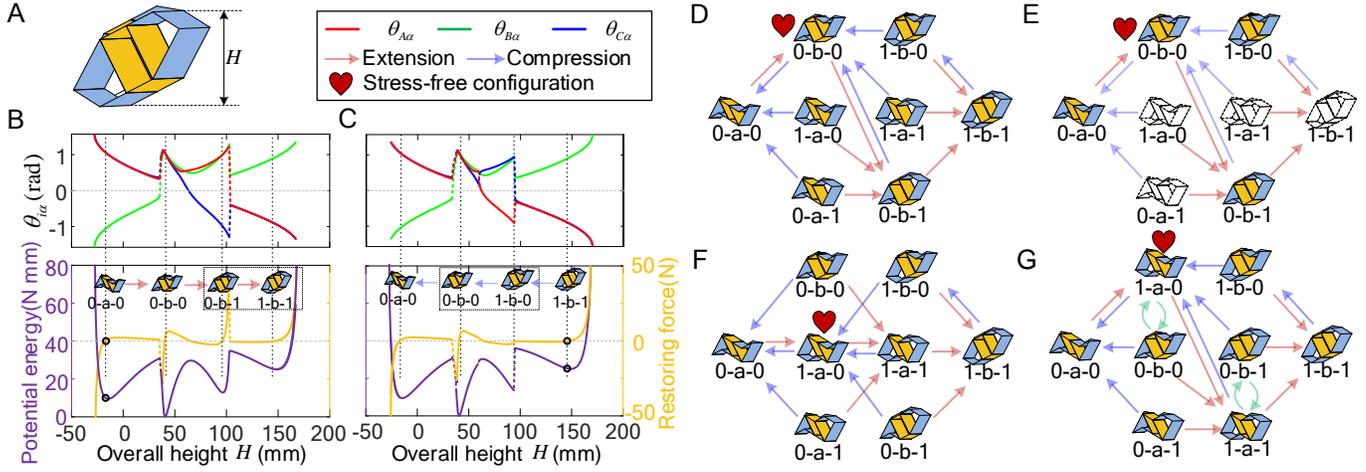

**Figure 2.** Quasi-static transition sequences of the SMOV cell under the displacement control with design parameters located at P1 in Fig. 1F and with equivalent stiffnesses $k_1^* = 500 k_\alpha$, $k_2^* = 700 k_\alpha$. (A) Illustration of the overall height of the SMOV cell. The evolutions of the folding angles of the constituent units A, B, C, the corresponding potential energy, and the restoring force with respect to the overall height with different initial configurations '0-a-0' and '1-b-1' are presented in (B) and (C), respectively. The black circles and the dotted lines denote the initial configurations and the stable configurations, respectively. The dashed box represents the irreversible transition. The whole transition map obtained by integrating the transition sequences starting from the 8 different initial configurations is shown in (D). Configuration marked with heart shape is stress-free. The transition map in (E) presents the situation with only 4 stable configurations with design parameters located at P2 in Fig 1F (configuration with white color is unstable). (F) Transition map with stress-free configurations '1-a-0'. (G) is the same map as (F) but with rearranged positions of the 8 stable configurations. Green arrows denote the configurations which changed their positions in the map.

It is worth pointing out that the minimum-energy path search, which is fundamentally an optimization process, closely relates to the loading direction as well as the initial configurations. As a result, to acquire a thorough understanding of the possible transition sequences, displacement controls (including extensions and compressions) starting from different stable configurations are applied to the SMOV cell. For example, with '0-a-0' as the initial configuration and by decreasing the overall height of the SMOV cell (*Fig. 2A*), i.e., compression, the potential energy, and the restoring force will increase sharply (*Fig. 2B*), while the cell will be folded to a flat state ($|\theta_{i\alpha}| \to \pi/2$ ($i = A, B, C$)) without any phase transition. On the contrary, by increasing the height of the SMOV cell from '0-a-0', i.e., extension, three configuration transitions to '0-b-0', '0-b-1', and '1-b-1' are identified via the optimization process, giving rise to a potential energy curve with four wells. Particularly, during the transitions from '0-a-0' to '0-b-0' and '0-b-1' to '1-b-1', the potential energy and the corresponding restoring force experience a discontinuous jump, manifested as a snap-through transition (see the jumps occurred on the folding angles of the constituent units, *Fig. 2B*, top). With the final configuration '1-b-1' as the starting point and by reversing the loading direction, i.e., compressing, a similar four-well potential curve and snap-through transitions are witnessed, while the stable configurations are no longer identical to those in the extension process. The SMOV cell will travel through a new stable configuration '1-b-0', which indicates that the transitions from '0-b-1' to '1-b-1' and from '1-b-0' to '0-b-0' are uni-directional and irreversible.

The transition map of the SMOV cell (*Fig. 2D*) can be obtained by integrating the transition sequences starting from the eight different initial configurations (*SI Appendix, Fig. S3*). Note that the map is not fully connected, instead, it is made up of uni-directional and bi-directional transitions. By tailoring the design parameters, the reversibility and irreversibility of the transition branches can be changed accordingly, giving rise to qualitatively different transition maps (*SI Appendix, Fig. S4*).

Recall that the stiffness ratio and the stress-free angle play a key role in determining the number of stable configurations. To understand how they affect the transition behavior of the SMOV cell, the transition maps corresponding to point P1 (with eight stable configurations) and point P2 (with four stable configurations) in *Fig. 1F* are illustrated in *Fig. 2D and 2E*, respectively. It reveals that in the transition map corresponding to point P2, configurations '1-a-0', '1-a-1', '1-b-1', and '0-a-1' are unstable (denoted by blank shapes with dashed edges), while the transition paths in the two maps are still identical. This suggests that the transition paths are not sensitive to the stiffness ratio and the stress-free angle. However, the stress-free configuration, which has been shown to be nonessential to the number of stable configurations, shows its capability to alter the transition sequences of the SMOV cell. For example, by switching the stress-free configuration from '0-b-0' to '1-a-0', the overall transition map is qualitatively changed (*Fig. 2F*): some reversible transitions become irreversible (e.g., between '0-b-0' and '0-a-0'), while some irreversible paths become reversible (e.g., between '0-a-0' and '1-a-0'); moreover, some new transition paths emerge in the new map (e.g., from '1-b-0' to '1-a-0').

Note that configurations '0-b-0' and '1-a-0' correspond to an almost identical height of the SMOV cell, so are configurations '0-b-1' and '1-a-1'. By exchanging the position of '0-b-0' with '1-a-0', and the position of '0-b-1' with '1-a-1' on the map (*Fig. 2G*), the transition paths could remain unchanged as those in Fig. 2D. A similar phenomenon is also observed in the case where configuration '0-a-1' serves as the stress-free configuration (*SI Appendix, Fig. S4C*). This can be interpreted from the fact that the transition paths are mainly determined by the overall height and the potential energy level of the SMOV cell. We further examine all the cases with the eight configurations serving as the stress-free states (*SI Appendix, Fig. S4*), a generic conclusion can be drawn. If the stress-free configurations are of different heights, the relative potential energy relationship among the nodes of the map is changed, and the generated transition maps are fundamentally different; while if the stress-free configurations are of almost identical height (e.g., *SI Appendix, Fig. S4A~C, and Fig. S4D~F*), by exchanging the designated configurations with similar height on the map, the relative height and energy relationship among the nodes of the map are retained, thus preserving the transition paths.

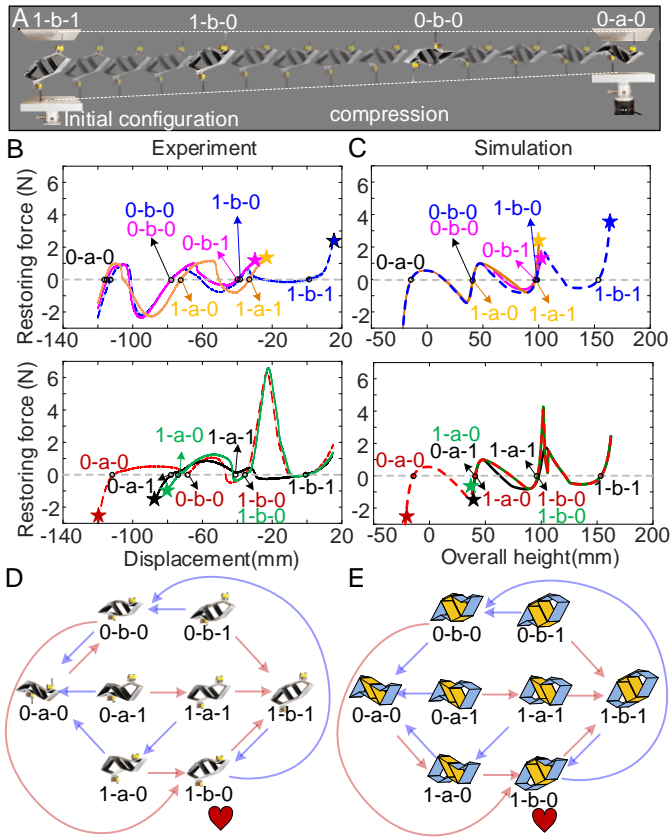

**Figure 3.** Experimental investigation of the quasi-static transition sequences. (A) Time-lapse photo of the SMOV prototype during a quasi-static compression test. (B) and (C) respectively show the experimental and numerical curves of the restoring force with respect to the external control height. (D) and (E) are the corresponding transition maps.

The transition map, consisting of reversible and irreversible transition paths, is also obtained via experiments on a SMOV prototype (see detailed fabrication and experimental setup in *SI Appendix, Section S5*). With '0-b-1' as the stress-free configuration and '1-b-1' as the initial configuration, by applying compressing displacement control, a series of configuration switches are observed, shown in the time-lapse photo (*Fig. 3A*) and the corresponding force-displacement curve (*Fig. 3B* top, in blue color). Note that the points on the curve with zero restoring force correspond to the stable configurations, '1-b-1', '0-b-1', '0-b-0', and '0-a-0', which constitute a chain of transition sequences. With the other stable configurations as the initial states and by applying extension/compression displacement control, different transition sequences can be achieved (*Fig. 3B and SI Appendix, Video S1*). Integrating these sequences together, the complete transition map can be generated (*Fig. 3D*).

Accordingly, based on the model with imperfect constraints and the optimization scheme, the transitions can also be obtained via numerical analysis (*Fig. 3C*), which agrees well with the experimental results (*Fig. 3B*) in terms of the number of stable configurations and the overall trend of the force-displacement curves. Quantitatively, the numerical and experimental results are also in good agreement. For example, both numerical simulations and experiments suggest that a small compression force is enough to trigger a snap-through transition from '1-b-1' to '0-b-1'(the blue dashed curve in *Fig. 3B and 3C, top*); while the required extension force for the reverse transition from '0-b-1' to '1-b-1' is much larger (the red dashed or green curves in *Fig. 3B and 3C, bottom*). Furthermore, comparing the transitions maps obtained from experiments (*Fig. 3D*) and simulations (*Fig. 3E*), we see that except for one transition, the two maps, consisting of uni-directional and bi-directional transition paths, exhibit convincing agreement with each other. This again manifests the effectiveness of the modeling and path-searching approaches.

**Dynamic transition**
In addition to quasi-static control, configuration transitions can be further enriched when the SMOV cell is subject to dynamic excitations. By considering the inertial effect of the facets and based on the Lagrange equation, the dynamic governing equation of the SMOV cell is derived and shown in *SI Appendix, Section S6*. When performing the dynamic simulation, the initial state is set at one of the stable configurations with zero velocity, and the excitation amplitude and frequency are swept. The steady-state response types, in terms of the equilibrium that the system oscillates around, are recorded and showed in a dynamic transition map (*Fig. 4A*), in which the configuration '0-b-1' is set as the initial state. Note that the transitions of the steady-state responses are closely related to the excitations. With relatively small excitation amplitude, the SMOV cell keeps oscillating around the '0-b-1' configuration without change; while with larger excitation amplitudes or higher excitation frequency (i.e., sufficiently high input energy), inter-well oscillations around multiple equilibria will be triggered. In the intermediate region, rich transitions of the steady-state responses are observed, with the surrounded stable equilibrium changing from '0-b-1' to the other seven stable configurations. Similar trends are also witnessed when the other seven stable configurations are set as the initial states (*SI Appendix, Fig. S7*). Being different from the quasi-static scenario in that certain stable configurations cannot be reversibly transformed, here, steady-state oscillations around any of the two stable configurations can be reversibly switched by applying proper dynamic excitations, generating a fully-connected dynamic transition map.

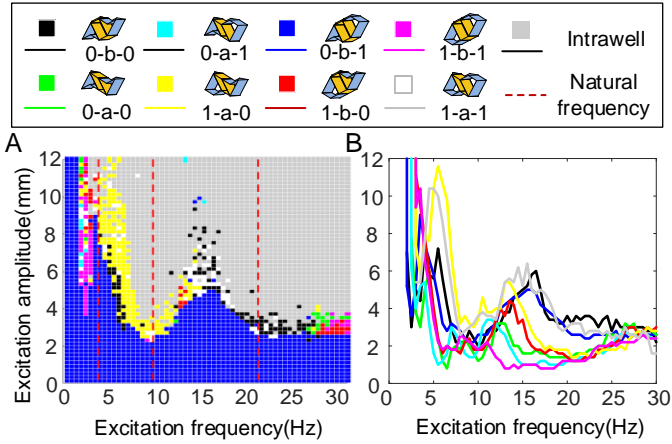

**Figure 4.** Transition sequences with periodic dynamic control. (A) Correlations between the dynamic transitions and the external excitations, i.e., excitation frequency and amplitude. The initial state is at the '0-b-1' configuration with zero initial velocity, and the dashed lines are natural frequencies of the linearized model. (B) critical lines triggering transitions with different initial configurations.

The dynamic transition map can be further interpreted from the perspective of resonance. To this end, the dynamic system is linearized around its stable equilibria such that the natural frequencies can be derived (see *SI Appendix, Table. S2*). In the eight cases shown in *Fig. 4A* and *Fig. S7*, the first three natural frequencies are denoted by red dashed lines. It reveals that around the natural frequencies, the required excitation amplitudes for transitions are obviously lower than those in other frequency ranges. This is because the response amplitude will be amplified significantly due to the resonance effect, which would thus overcome the energy barrier between adjacent stable equilibria and induce a snap-through motion.

To quantify the required energy level for triggering dynamic transitions, the critical curve on each map is extracted, and they are depicted in *Fig. 4B*. Starting from a certain initial state, if the excitation condition locates below the corresponding critical curve, the SMOV cell will keep its intra-well oscillation around the initial stable state. Above the critical curve but below those corresponding to the other initial stable configurations, the stable equilibrium that the steady-state oscillation surrounds are available to change. When the excitation condition locates above all critical curves, large-amplitude inter-well oscillation will take place.

**Transition sequences for mechano-logic**
As discussed, under quasi-static loading or dynamic excitations, the SMOV cell could exhibit rich transition behaviors. The transition maps with reversible/irreversible paths are promising in many applications, such as reconfigurable robots and reprogrammable metamaterials. In this research, we especially showcase a novel and unique potential of the SMOV in achieving mechano-logic, which endows computing ability in the mechanical domain.

First of all, the SMOV cell is capable of realizing the functionality of the basic logic gates, i.e., AND, OR, and XOR gates (*Fig. 5A*). Note that digital inputs '0' or '1' are the objects that these logical operation functions will process. Hence, the SMOV cell's stable configurations are encoded. Specifically, for units A and C of the SMOV cell, as the previously used denotation, the 'bulged-out' and the 'nested-in' stable configurations are respectively converted into digits '1' and '0'; for unit B, the 'inclined-up' and the 'inclined-down' stable configurations are respectively put into '1' and '0'. With such encoding, the configurations of the SMOV cell can be represented by three digits (e.g., the initial configuration is assigned to be '0-0-0'). Without loss of generality, the digits of units A and B are specified as the input of the logic gate. The logic operation is achieved by state transitions under a prescribed control, which can be a quasi-static displacement loading or a dynamic excitation. Here, a quasi-static extension process (with only one control step) is employed to exemplify the logic operation. For different inputs, the stable configuration of the SMOV cell will be transformed, following the transition map obtained in *Fig. 3D*. By recording the digits of the transited stable configurations, an augmented matrix $X$ is constructed (*Fig. 5A*). In addition to the inputs represented in the first two columns of the matrix $X$, more columns obtained by state transitions are included. The augmented matrix, which is fundamentally the spatial-temporal pattern of a physical reservoir, provides rich possibilities for complex logic operation. The output of the logic operation is achieved by a linear readout layer with weights $W_{out}$, i.e., the output $\hat{Y} = sigmoid(X \cdot W_{out})$. By optimizing the linear weights, the three logic gates (AND, OR, and XOR) can be successfully realized based on the same SMOV cell (*Fig. 5A*, bottom).

To understand the importance of state transitions in achieving the logic operations, a direct readout framework without state transitions is illustrated as a comparison (*Fig. 5B*, top). It shows that direct readout from the initial configurations of the SMOV cell could successfully realize the AND and OR gates, however, fail in the XOR operation (*Fig. 5B*, bottom). This is because for the AND and OR gates, the mapping from the input onto the output is essentially linear, which, as a result, could be distinguished via a direct linear readout classifier. However, the input-output mapping for the XOR gate is nonlinear, and a linear classifier would be inapplicable. On the contrary, by introducing the state transitions (i.e., the physical reservoir, achieved by quasi-static extension) into the operation process (*Fig. 5A*), the initial configurations (i.e., the inputs) can be transited to other stable states, which, fundamentally, is a nonlinear transformation in terms of the digits. As a result, different logic gates, including the ones with nonlinear input-output mapping, can be realized.

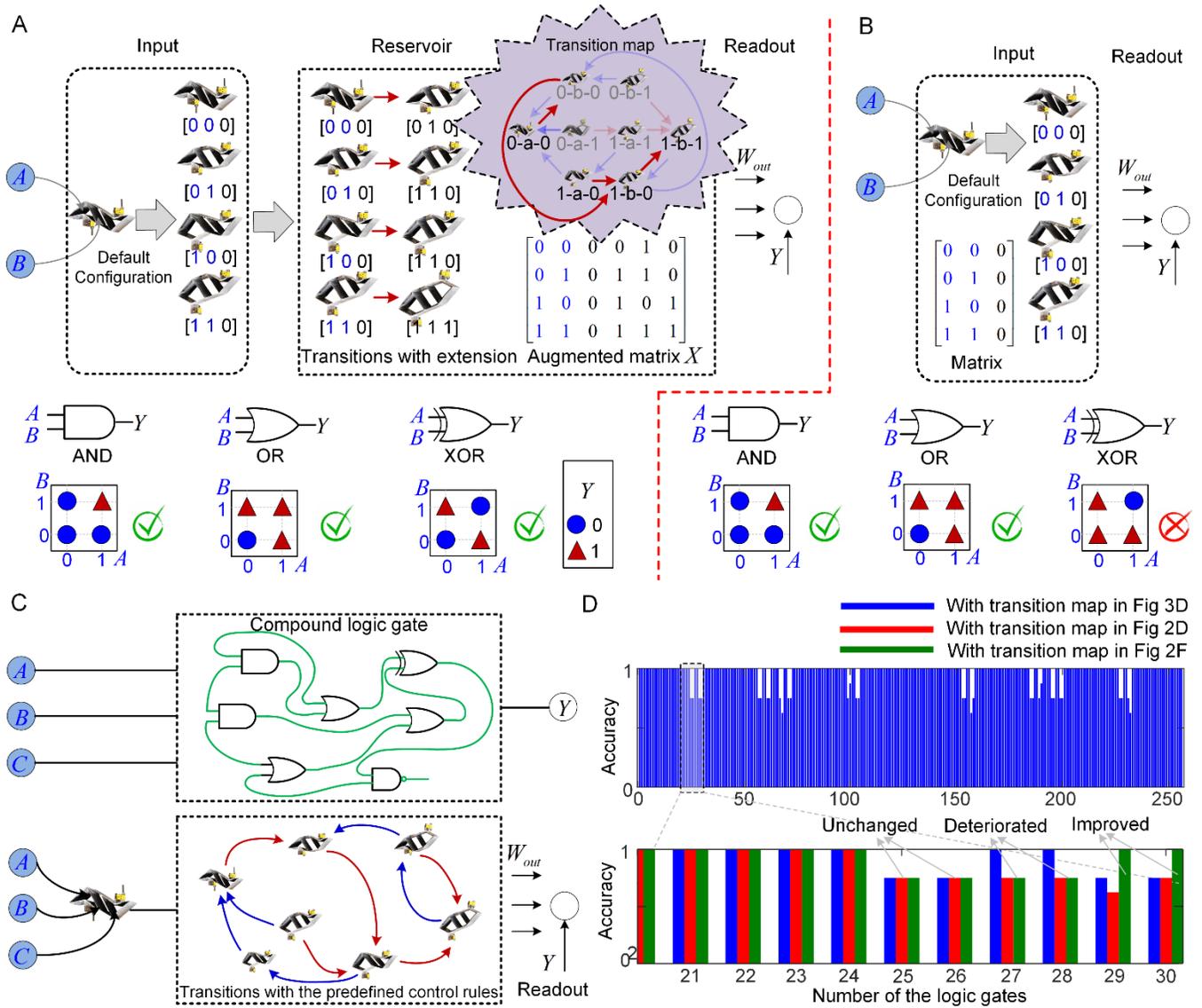

**Figure 5.** The architecture and results of using the SMOV cell to develop basic logic operations (A) with transition procedures, and (B) without transition procedures. (C) The conceptual framework of developing compound logic gates with three inputs and one output in the electrical scheme and our proposed mechanical scheme. (D) Prediction accuracy of using the SMOV cell with different transition maps for all compound logic gates with three inputs and one output. Top: with the experimental transition map; Bottom: comparisons among the three transition maps.

More complex logic gates, such as a compound logic shown in SI Appendix, *Fig. S8A* can also be achieved based on the SMOV cell by adding another input unit and incorporating more control steps. Specifically, the fuzzy computing in a compound logic gate, manifesting as connections of multiple AND gates, is equivalently achieved by conducting two transitions steps following the transition map and a subsequent linear readout procedure with weights $W_{out}$. Unlike the conventional mechano-logic approach that multiple basic logic gates have to be integrated to realize compound logic, our scheme can achieve complex logic operations based on a single SMOV cell, without increasing the structural complexity. This merit originates from the transition behaviors of the multistable SMOV cell, which is a reflection of the nonlinearity of the physical reservoir. As a further example, we demonstrate that a full adder, which is central to most digital circuits that perform addition or subtraction, can also be developed based on a single SMOV cell, see *SI Appendix, Fig. S8B*. The three inputs of a full adder are the operands A, B, and the input carry $C_{in}$; the output of a full adder is the final sum output S and the final carry output $C_{out}$. With three inputs and two outputs, two transition steps are needed, and an additional set of readout weights ($W_{out}^*$) is incorporated. Note that the versatility of the SMOV cell for different logic operations is closely associated with the readout. The weights of the readout layer are trainable by analyzing the spatial-temporal patterns of the reservoir so that a single SMOV cell is capable of achieving different logic operations.

Dynamic transitions of the SMOV cell's configurations, which have been shown to be richer than the quasi-static transitions, can also serve as the physical reservoir for logic operations. The difference lies in that a sinusoidal excitation, instead of quasi-static displacement control, is applied to generate the transition sequences (*SI Appendix, Fig. S8C*). With dynamic transition sequences and the associated readout, the SMOV cell can also perform the basic logical operations (i.e., AND, OR, and XOR).

Note that for a compound logic gate with three inputs and one output (see a conceptual example in *Fig. 5C*, top), $2^8$ different input-output mappings are possible, which correspond to $2^8$ logic gates. They can be equivalently achieved via the transitions of the SMOV cell and the trained readout (*Fig. 5C*, bottom). Note that not all the $2^8$ logic gates can be accurately achieved. For example, by utilizing the experimental transition map (*Fig. 3D*) and with two transition steps in the proposed scheme, the prediction accuracy of the $2^8$ logic gates is tested and shown in *Fig. 5D* top. Note that if the accuracy is lower than 100%, the corresponding logic gate cannot be realized. Such cases are indeed observed in *Fig. 5D* (Top). With different transition maps (e.g., the maps shown in *Fig. 2D* and *Fig. 2F*) obtained by adjusting the design parameters of the SMOV cell, the prediction accuracy would be modified. Some of the logic gates that cannot be realized via the experimental transition map are now achievable via another transition map (*Fig. 5D*, bottom). Actually, with different designs of the SMOV cell and different control rules, distinct transition behaviors (i.e., the reservoir) can be obtained, which could be tailored for different mechano-logic. Particularly, if dynamic excitations are used as the control strategy, the fully-connected dynamic transition map could further improve the SMOV logic operations.

## Conclusion

A Miura-variant metamaterial with exceptional multistability and reconfigurable features has been leveraged and investigated as a platform to uncover the deep knowledge and understanding of harnessing multistability transition sequences in both the quasi-static and dynamic realms. By introducing controllable flexibility into the SMOV cell and via a combination of theoretical, numerical, and experimental efforts, rich transition sequences that are predictable and discriminative, including reversible and irreversible paths, are revealed. In addition, the underlying mechanism for editing the transition maps via tailoring the design parameters is uncovered. Dynamic excitations can also trigger transitions, manifested as steady-state oscillations around different stable states. Different from the quasi-static scenario, bi-directional dynamic transitions are accessible between any of the two stable configurations, which constitute a fully-connected transition map. Insights into triggering the transitions are obtained in terms of the resonant frequency and the injected energy.

The SMOV cell, as a representative multistable structure, provides a new path for developing mechano-logic. Based on a single SMOV cell and by harnessing the quasi-static/dynamic transitions as a physical reservoir, basic and complex logical operations are achieved. The proposed framework endows the SMOV cell with the versatility of using one structural element to conduct different logic operations, which greatly reduces the complexity for developing various compound logic gates. Such merit originates from the nonlinearity of the multistable transition. Benefiting from this and by constructing a multi-cell SMOV metamaterial, it is promising in achieving complex computing. Building upon this foundation, we also expect that rich intelligent functionalities, such as sensing, signal processing, decision making, and control, which at present rely solely on the computation power of the electronic components, can be embedded and realized in the mechanical domain and greatly inspire the creation of materials with physical intelligence.

## Materials and Methods

Derivations of the geometry of the rigid-foldable origami are presented in SI Appendix, sections S1 and S2. For the scenario that the rigid-foldability cannot be ideally satisfied, the modeling and the optimization method for deriving the quasi-static transitions are introduced in SI Appendix, sections S3 and S4. Materials and fabrication methods of the experimental prototype are summarized in SI Appendix, section S5. Sections S6 and S7 of the SI Appendix detail the dynamic modeling and dynamic transition sequences, respectively.

## Acknowledgments

Z.L., H.F., and J.X. acknowledge the supports from the National Key Research and Development Project of China under Grant No. 2020YFB1312900 and the Key Project of the National Natural Science Foundation of China under Grant No. 11932015. Z.L. also acknowledges the China Postdoctoral Science Foundation under Grant No. 2021TQ0071 and 2021M700819, and H.F. acknowledges the National Natural Science Foundation of China under Grant No. 11902078. This research is also partially supported by the University of Michigan Collegiate Professorship.

**Supplementary Information Text**

**S1. Design and kinematics of a unit cell**

The construction of the multistable origami metamaterials has been introduced in the main text. Here we detail the geometry and kinematics. The metamaterials can be fabricated by two different Miura-ori sheets, denoted by $\alpha$ and $\beta$, see Fig. S1*A*. The acute angles and crease lengths of the two sheets are $\gamma_\alpha$, $\gamma_\beta$, and $a_\alpha$, $b_\alpha$, $a_\beta$, $b_\beta$, respectively. The geometry compatibility for stacking the two sheets is expressed as

$$b_\alpha = b_\beta = b, \frac{\cos \gamma_\alpha}{\cos \gamma_\beta} = \frac{a_\beta}{a_\alpha}. \quad [1]$$

The units show different configurations in Fig. 1*D* of the main text; however, the inner kinematics of the units are similar. Both of the two different units are a single degree of freedom system considering the rigid-folding assumption. Therefore, the dihedral angles in each unit can be solely described by $\theta_{A\alpha}$, $\theta_{B\alpha}$ and $\theta_{C\alpha}$, respectively, which are dihedral angles between the sheet $\alpha$ and the $x-y$ plane in units A, B, and C, respectively. Specifically, units A and C share the same geometry and kinematics. The dihedral angles between sheet $\beta$ and $x-y$ plane can be derived as

$$\theta_{i\beta} = \arccos\left(\cos\theta_{i\alpha} \frac{\tan \gamma_\alpha}{\tan \gamma_\beta}\right), \quad [2]$$

where $i$ stands for A and C. Other dihedral angles can all be described as a function of $\theta_{i\alpha}$ and $\theta_{i\beta}$:

$$\rho_{i7} = \rho_{i8} = \pi - 2\theta_{i\alpha}, \rho_{i3} = \rho_{i4} = \pi - 2\theta_{i\beta}, \rho_{i5} = 2\arccos\left(\frac{\sin\theta_{i\alpha}\cos\gamma_\alpha}{\sqrt{1-\sin^2\theta_{i\alpha}\sin^2\gamma_\alpha}}\right),$$

$$\rho_{i1} = 2\arccos\left(\frac{\sin\theta_{i\beta}\cos\gamma_\beta}{\sqrt{1-\sin^2\theta_{i\beta}\sin^2\gamma_\beta}}\right), \rho_{i6} = 2\pi - \rho_{i5}, \rho_{i2} = 2\pi - \rho_{i1}, \quad [3]$$

and the dihedral angle at the connecting crease is $\rho_{i0} = \theta_{i\beta} - \theta_{i\alpha}$. The heights of each constituent Miura-ori sheet $\alpha$ and $\beta$ are

$$H_{i\alpha} = -a_\alpha \sin\theta_{i\alpha}\sin\gamma_\alpha, H_{i\beta} = a_\beta \sin\theta_{i\beta}\sin\gamma_\beta. \quad [4]$$

Therefore, the overall height of the unit is $H_k = H_{i\alpha} + H_{i\beta}$. Besides, the length and width yield

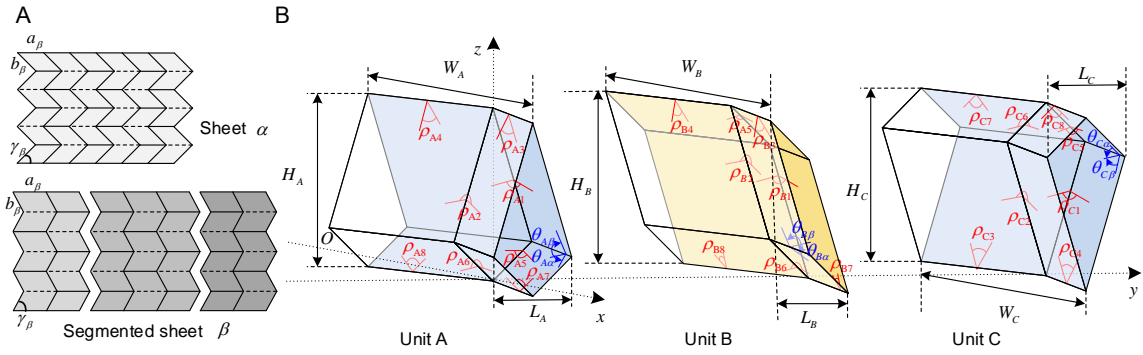

**Fig. S1** Geometry and kinematics of the constituent units A, B, and C of a single cell. The dihedral angles and the external dimensions of each unit are denoted.

$$L_i = 2a_\alpha \sqrt{1 - \sin^2 \theta_{i\alpha} \sin^2 \gamma_\alpha}, \quad W_i = \frac{2b \cos \theta_{i\alpha} \tan \gamma_\alpha}{\sqrt{1 + \cos^2 \theta_{i\alpha} \tan^2 \gamma_\alpha}}. \quad [5]$$

$\theta_{B\beta}$, the dihedral angle between sheet $\beta$ and $x-y$ plane in unit B, is the supplementary angle of $\theta_{B\alpha}$, i.e. $\theta_{B\alpha} = \pi - \theta_{B\alpha}$. Similarly, the other dihedral angles in unit B can be derived as

$$\rho_{B5} = 2\arccos\left(\frac{\sin\theta_{B\alpha}\cos\gamma_\alpha}{\sqrt{1-\sin^2\theta_{B\alpha}\sin^2\gamma_\alpha}}\right), \rho_{B1} = 2\arccos\left(\frac{\sin\theta_{B\beta}\cos\gamma_\beta}{\sqrt{1-\sin^2\theta_{B\beta}\sin^2\gamma_\beta}}\right),$$
$$\rho_{B6} = 2\pi - \rho_{B5}, \rho_{B2} = 2\pi - \rho_{B1}, \rho_{B0} = \theta_{B\beta} - \theta_{B\alpha},$$
$$\rho_{B7} = \rho_{B8} = \rho_{B3} = \rho_{B4} = \pi - \rho_{B0}, \quad [6]$$

and the outer dimensions are

$$L_B = 2a_\alpha \sqrt{1 - \sin^2 \theta_{B\alpha} \sin^2 \gamma_\alpha}, \quad W_B = \frac{2b \cos \theta_{B\alpha} \tan \gamma_\alpha}{\sqrt{1 + \cos^2 \theta_{B\alpha} \tan^2 \gamma_\alpha}}, \quad H_B = H_{B\alpha} + H_{B\beta}, \quad [7]$$

where

$$H_{B\alpha} = -a_\alpha \sin\theta_{B\alpha} \sin\gamma_\alpha, \quad H_{B\beta} = a_\beta \sin\theta_{B\beta} \sin\gamma_\beta. \quad [8]$$

## S2. Parameter study and multistability

If the rigid-folding kinematics are strictly followed, i.e. the facets are rigid and the creases are ideal hinge-like, the unit cell degenerates into a single degree of freedom system, that is, $\theta_{A\alpha} = \pm\theta_{B\alpha} = \pm\theta_{C\alpha}$. Different torsional spring stiffness per unit length, $k_\alpha$ and $k_\beta$, are assigned to the folding creases of sheets $\alpha$ and $\beta$, respectively. In addition, the torsional spring stiffness per unit length at the connecting creases is assumed to be the same as that in sheet $\alpha$. Therefore, the total potential energy of the SMOV cell is the summation of the potential energy of the three units

$$\Pi_{ABC} = \frac{1}{2} \sum_{J=A,B,C} \left[ \sum_{i=1}^{8} K_{Ji} \left( \rho_{Ji} - \rho_{Ji}^0 \right)^2 + 4K_{J0} \left( \rho_{J0} - \rho_{J0}^0 \right)^2 \right], \qquad [9]$$

where $K_{Ji}$ is the torsional spring stiffness corresponding to the dihedral angle $\rho_{Ji}$, which is the product of the corresponding torsional spring stiffness per unit length and the crease length. $\rho_{Ji}^0$ is the dihedral angle of the stress-free configuration, where no internal forces exist in the SMOV cell. Given the stress-free angle $\theta_{A\alpha}^0$ of unit A, and the stress-free configurations relating to the kinematic relationship $\theta_{A\alpha}^0 = \pm\theta_{B\alpha}^0 = \pm\theta_{C\alpha}^0$, the stress-free dihedral angle $\rho_{Ji}^0$ can be obtained by substituting $\theta_{A\alpha}^0$, $\theta_{B\alpha}^0$ and $\theta_{C\alpha}^0$ into Eqs. [2], [3], and [6].

In the paper, the dimensions of the Miura-ori sheets are set as $b = a_\alpha = 38.1\text{mm}$, $\gamma_\alpha = 60°$ and $\gamma_\beta = 75°$. Let $k_\beta = \mu k_\alpha$, where $\mu$ is the stiffness ratio of the two sheets. Without loss of generality, we set $k_\alpha = 0.01\text{N/rad}$. We show that three design parameters, i.e., the stiffness ratio, stress-free angles, and stress-free configurations, can significantly alter the multistability of the SMOV cell. Specifically, the SMOV cell with different design parameters may have different number of stable configurations or the same number but with different specific shapes of the stable configurations.

We set the stress-free angle at $\theta_{A\alpha}^0 = \pi/3$, and the stress-free configuration at '0-b-0'. Let $\partial\Pi_{ABC}/\partial\theta_{A\alpha} = 0$, the local minimum and local maximum in the potential energy landscape, which correspond to the stable and unstable configurations, are derived. In Fig. S2*A*, we show the variation trend of the folding angle $\theta_{A\alpha}$ corresponding to the stable and unstable configurations concerning the changing of the stiffness ratio. Note that configuration with $\theta_{A\alpha} = \pi/3$ is always stable because it corresponds to the stress-free angle. However, the number of stable configurations and the corresponding folding angles $\theta_{A\alpha}$ vary with the stiffness ratio. Specifically, with a small stiffness ratio (the region Ⅰ in Fig. S2*A*), only four configurations are stable, and no unstable configurations exist. The detailed potential energy landscape in this region is illustrated in Fig. 1*H* top with $\mu=5$, which shows the mono-stability of the unit with each kinematic constraint. Then, with a larger stiffness ratio (region Ⅱ), two new configurations '0-a-1' and '1-a-0' become stable. Because of the symmetry, there two configurations always become stable simultaneously. Fig. S2*C* shows how the potential energy profile evolves with $\mu=6.5$. Continue increasing the stiffness ratio, new potential energy well emerges in configuration with constraints $\theta_{A\alpha} = -\theta_{B\alpha} = \theta_{C\alpha}$, which make the configuration '1-b-1' stable, see Fig. S2D and the corresponding region III in Fig. S2A. All 8 configurations become stable once the stiffness ratio is sufficiently large in the region IV. Exemplified potential energy landscape with 8 wells is presented in Fig. 1*G* top.

Similarly, we fix the stiffness ratio at $\mu=20$ and the stress-free configuration at '0-b-0' to investigate the influence exerted by the stress-free angle. The variation of the folding angle $\theta_{A\alpha}$ corresponding to the stable configurations concerning the change of $\theta_{A\alpha}^0$ is illustrated in Fig. S2B. Note that all the lines intersect at $\theta_{A\alpha}^0 = 0^o$, which corresponds to the kinematic bifurcation point with only one stable configuration. Except that, there are still four different regions corresponding to a different number of stable configurations. Similarly, it can be concluded that more stable configurations emerge when the stress-free angle deviates away from the critical condition.

The combined effect of the stiffness ratio and the stress-free angle on the multistability of the unit cell is illustrated in Fig. 1*F*. Besides, the effects of stress-free configurations on the potential energy landscape are illustrated in Fig. 1*G* and *H*.

To conclude, the number of stable configurations of a unit cell is determined by the stress-free angle and the stiffness ratio. More stable configurations are generated as the stiffness ratio becomes sufficiently large, and the stress-free angle deviates further away from $0^o$. The stress-free configuration does not influence the number of stable configurations but shows the ability to change the specific shapes of the stable configurations. These three parameters are capable of tuning the multistability of the unit cell in different ways, which is critical in determining the transition sequences.

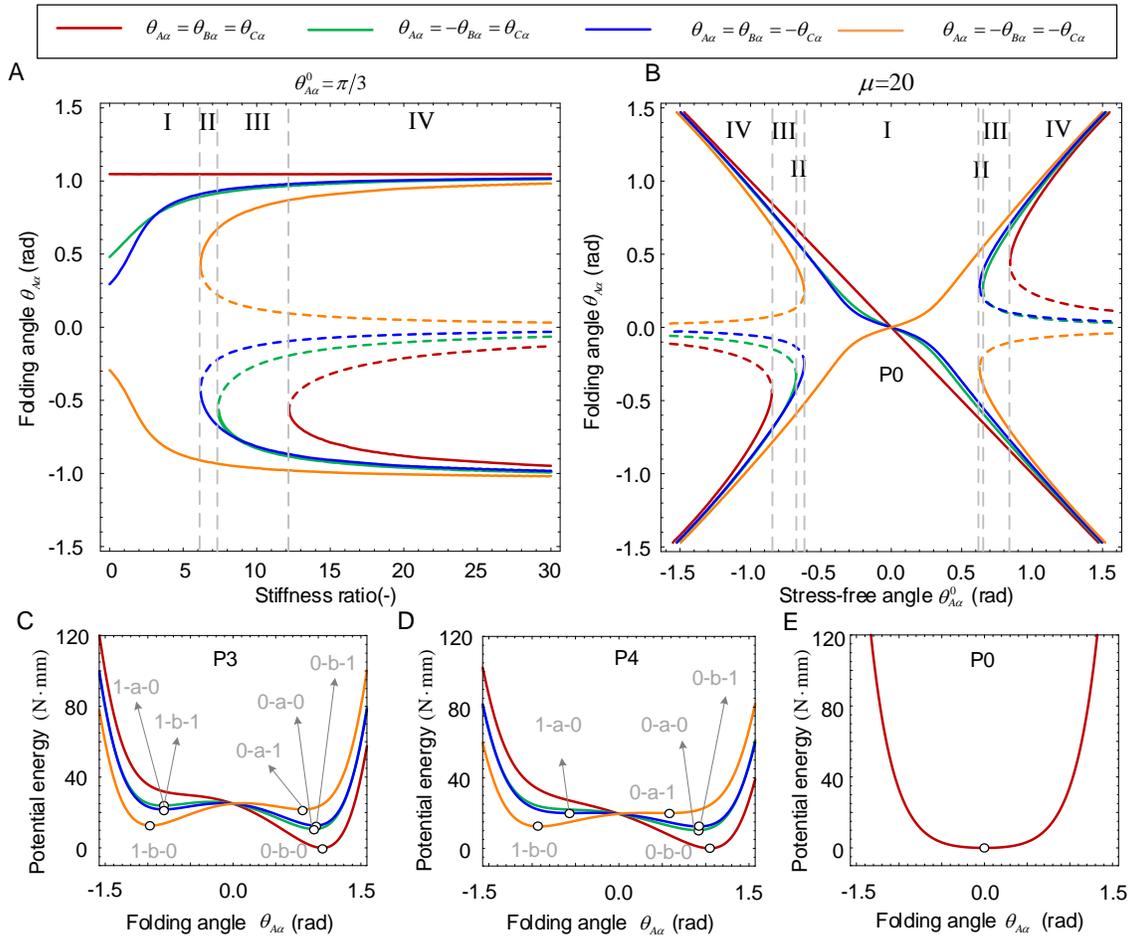

**Fig. S2** Multistability analysis of a single SMOV cell. The variation trend of folding angle $\theta_{A\alpha}$ corresponding to the stable and unstable configurations concerning the changing of (A) the stiffness-ratio, (B) the stress-free angle. (C) and (D) are the potential energy of the SMOV cell concerning the folding angle with the design parameters located at points P3 and P4 in Fig. 1*F*. (E) is the critical potential energy landscape with zero stress-free angles.

**S3. Potential energy with non-ideal constraints**

In the strict rigid-folding kinematics, the constraints $\theta_{A\alpha} = \pm\theta_{B\alpha} = \pm\theta_{C\alpha}$ force the three units to deform synchronously. A kinematic bifurcation point, i.e., $\theta_{A\alpha} = \theta_{B\alpha} = \theta_{C\alpha} = 0$, is encountered when transforming the SMOV cell among its stable configurations. After this point, the sign of the folding angle of each unit is not deterministic, making the transition sequences unpredictable. To tackle this issue, imperfect constraints originate from the flexibility of the structure are introduced in the design and fabrication process, which loosens the strict rigid-folding kinematic constraints by allowing each unit to deform asynchronously. However, the folding of adjacent two units is not fully independent; rather, the connecting surfaces still impose strong constraints to constrain the deviation between the folding of the adjacent cells. Here, two equivalent stiffness $k_1^*$ and $k_2^*$ are introduced to quantify these strong constraints. Specifically, the deviation between the dihedral angle $\rho_{A1}$ in unit A and the dihedral angle $\rho_{B1}$ in unit B, as well as the deviation between the dihedral angle $\rho_{B2}$ in unit B and the dihedral angle $\rho_{C2}$ in unit C, are utilized to uniquely characterize the folding deviation between the adjacent units, and will not be affected by the configuration discrepancy. Hence, the extra energy induced by these two extra constraints are given as

$$\Pi_{AB} = \frac{1}{2}k_1^*b(\rho_{A1} - \rho_{B1})^2, \Pi_{BC} = \frac{1}{2}k_2^*b(\rho_{B2} - \rho_{C2})^2. \qquad [10]$$

If the rigid-folding kinematics are strictly followed, no deviation exists between the adjacent units, i.e., $\rho_{A1} = \rho_{B1}$ and $\rho_{B2} = \rho_{C2}$. Therefore, extra energy is no longer applied. However, this extra energy gets higher with the increase of the deviation between the adjacent units.

The total potential energy of the imperfection modeling is the summation of the elastic energy induced by the folding of the hinge-like creases and by these extra constraints:

$$\Pi = \Pi_{ABC} + \Pi_{AB} + \Pi_{BC} = \frac{1}{2}\sum_{J=A,B,C}\left[\sum_{i=1}^{8}K_{Ji}(\rho_{Ji} - \rho_{Ji}^0)^2 + 4K_{J0}(\rho_{J0} - \rho_{J0}^0)^2\right]$$
$$+ \frac{1}{2}k_1^*b(\rho_{A1} - \rho_{B1})^2 + \frac{1}{2}k_2^*b(\rho_{B2} - \rho_{C2})^2. \qquad [11]$$

Particularly, if $k_1^* \to \infty$ and $k_2^* \to \infty$, the local minima of the total energy lies on $\rho_{A1} = \rho_{B1}$ and $\rho_{B2} = \rho_{C2}$, which degenerates the SMOV cell into a single-degree-of-freedom system. On the contrary, if $k_1^* \to 0$ and $k_2^* \to 0$, folding of the three units is completely independent. For the imperfect rigid-folding scenario, compromised constraints (large but not infinite stiffness) are assigned to characterize the practical constraints.

The local minima of the potential energy $\Pi_{ABC}$ with rigid-folding scenario has been thoroughly discussed in S2. At these local minima, constraints $\theta_{A\alpha} = \pm\theta_{B\alpha} = \pm\theta_{C\alpha}$ are always satisfied, indicating that no deviation (i.e., $\rho_{A1} = \rho_{B1}$ and $\rho_{B2} = \rho_{C2}$) exists between the adjacent units and no extra potential energy ($\Pi_{AB}$ and $\Pi_{AB}$) emerges at the stable configurations. Therefore, the potential energy landscape with the imperfection modeling is the same as the one with rigid-folding assumptions at these stable configurations. In this scenario, the multistability analysis with rigid folding assumption applies to the imperfection modeling. However, the landscape beyond the stable configurations is dramatically changed because of the introduction of the extra constraints, which in particular, determine the transition sequences.

**S4 quasi-static transition sequences**

The deterministic transition sequence can be obtained by optimizing Eq. [11] with given displacement control based on that the unit always evolves to the configuration in the neighborhood of its current state. Specifically, given a certain displacement $H_c$ and the last state $(\theta^c_{A\alpha}, \theta^c_{A\alpha}, \theta^c_{A\alpha})$, the corresponding configuration of the unit can be obtained by searching the minimum potential energy $\min(\Pi(\theta_{A\alpha}, \theta_{B\alpha}, \theta_{C\alpha}))$ in the neighborhood of $(\theta^c_{A\alpha}, \theta^c_{A\alpha}, \theta^c_{A\alpha})$ with the constraint $H(\theta_{A\alpha}, \theta_{B\alpha}, \theta_{C\alpha}) = H_c$. This local optimization with constraints is realized by the fmincon function in Matlab.

In the simulation, the equivalent stiffnesses are set as $k_1^* = 500k$, $k_1^* = 700k$, and the design parameters, i.e., the stiffness ratio, the stress-free angle, and the stress-free configuration, are adopted as $\mu=20$, $\theta^0_{A\alpha} = \pi/3$, and $\theta^0_{A\alpha} = \theta^0_{B\alpha} = \theta^0_{C\alpha}$, respectively. This set of parameters correspond to point P1 in Fig. 1*F*, where the SMOV cell has the most stable configurations (8 stable configurations). We systematically apply the extension and compression displacement control starting from every possible stable configuration to have a thorough overview of all the sequences. The corresponding potential energy profile, as well as the restoring force during the loading, are determined with the optimization method, see Fig. S3. The extracted transition sequences are illustrated in Fig. S4*A*. Similarly, transition maps with other 7 stress-free configurations are also derived in Fig. S4*B-H*.

By switching the equivalent stiffnesses, i.e., $k_1^* = 700k$, $k_2^* = 500k$, the corresponding transition maps have been changed (Fig. S4*I*), owing to the symmetry of the first and third units. We show that the structure will always follow the transition map in Fig. S4*A* when $k_1^* < k_2^*$, and follow the transition map in Fig. S4*I* when $k_1^* > k_2^*$, see Fig. S4*J*. Especially, when $k_1^* = k_2^*$, the first and the third unit become symmetry, therefore, some transitions become unpredictable again.

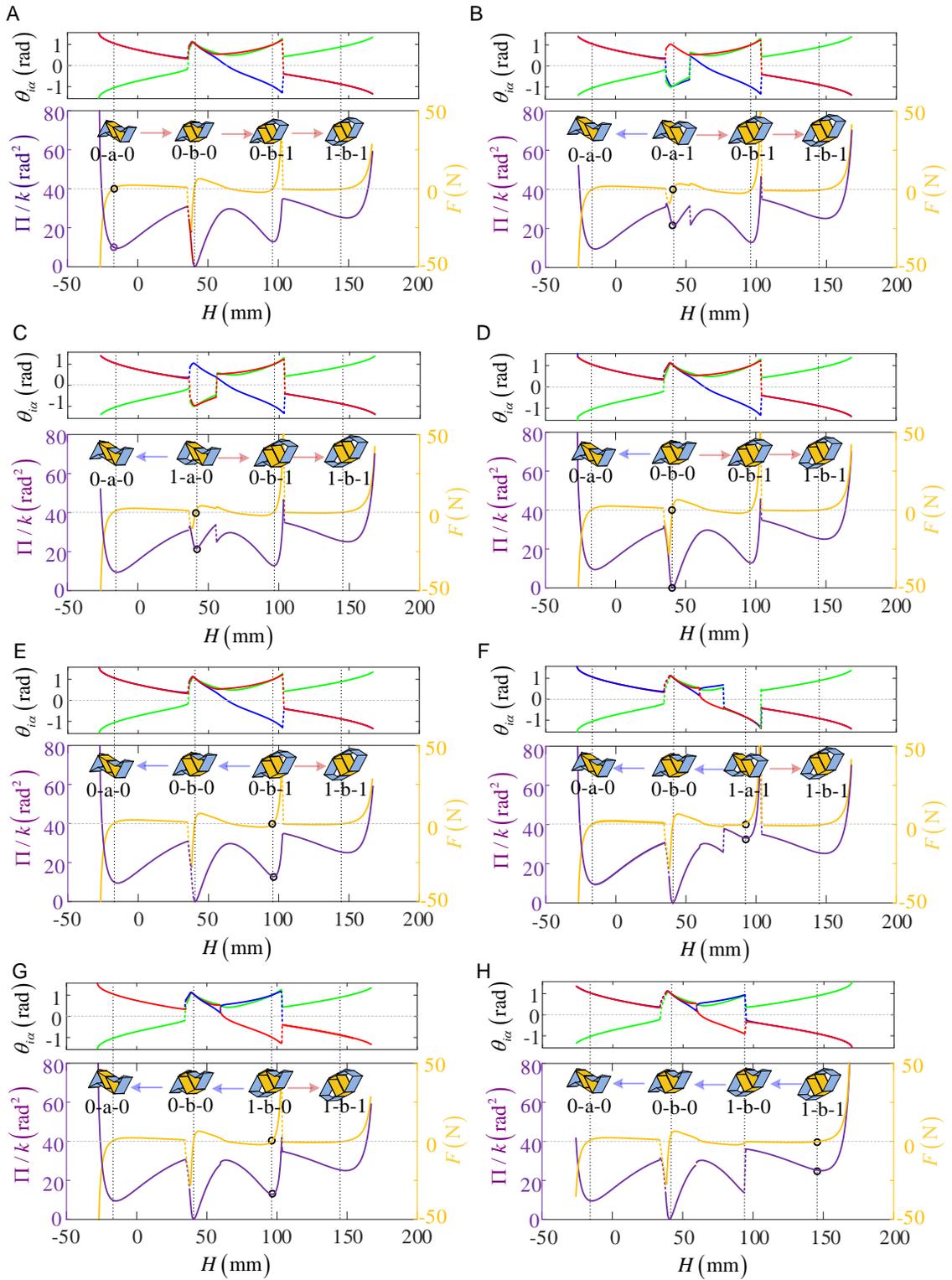

**Fig. S3** Folding angles, the corresponding potential energy landscape as well as the corresponding restoring force with respect to the overall height of the SMOV cell, starting from (A) '0-a-0', (B) '0-a-1', (C) '1-a-0', (D) '0-b-0', (E) '0-b-1', (F) '1-a-1', (G) '1-b-0', (H) '1-b-1' initial configurations.

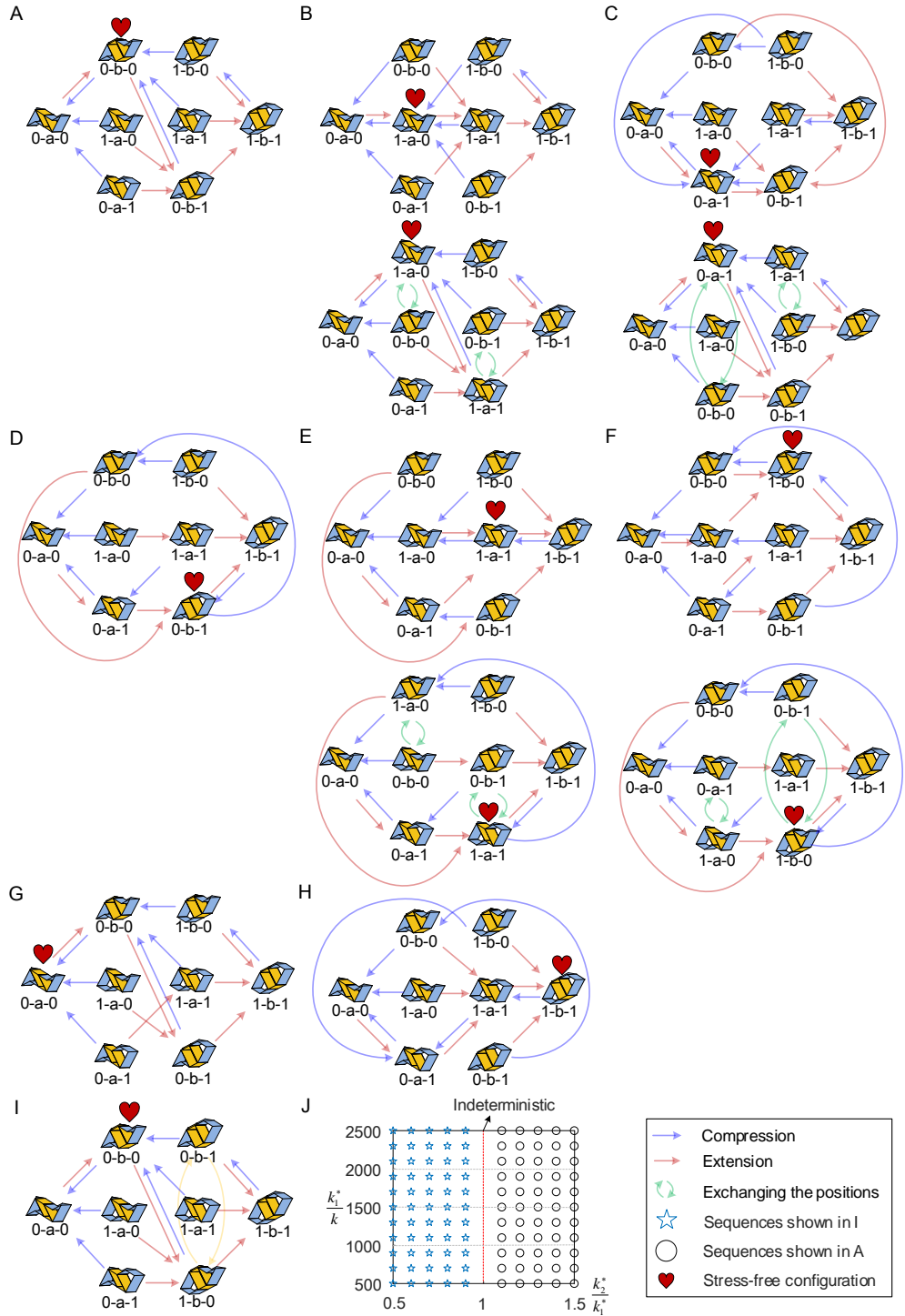

**Fig. S4** Transition sequences with quasi-static displacement control. The whole transition map of transition sequences with stress-free configuration (A) '0-b-0', (B) '1-a-0', (C) '0-a-1', (D) '1-b-1', (E) '0-b-1', (F) '1-a-1', (G) '1-b-0', (H) '0-a-0' and design parameters located at P1 in Fig. 1*F*. (I) The transition map with the switched value of $k_1^*$ and $k_2^*$. (J) Effects of $k_1^*$ and $k_2^*$ on the transition map. In the bottom plot of (B), (C), (E), and (F), positions of the 8 stable configurations are rearranged (denoted by green arrows) to reserve the map architecture.

## S5. Prototype and experiment setup

A proof-of-concept origami prototype is fabricated to verify the rich transition sequences. Specifically, the origami facets are water jet cut individually from 0.25-mm-thick stainless steel sheets. Then they are stuck to a 0.13-mm-thick adhesive-back plastic film [ultrahigh molecular weight (UHMW) polyethylene] to form the prescribed two different Miura-ori sheets, see Fig. S5*A*. After that, we fold the sheets in the way presented in Fig. S5*B*, and paste 0.01-mm-thick pre-bent spring-steel stripes at the corresponding creases to provide strong torsional stiffness. In this way, the stiffness ratio is greatly increased which will generate more stable configurations. The stress-free angle corresponding to a stress-free configuration is about $-\pi/3$. Then the sheets are connected along the connecting creases by adhesive films to form a complete single-cell prototype. Therefore, the stress-free configuration is '1-b-0', see Fig. S5*C*. In the experiment, we design a 3D-printed connector, which can be screwed onto the prototype with the rectangular steel plates. A screw rod is then utilized to connect the 3D-printed connector with the Instron machine. Starting from each different stable configuration, six complete extension and compression tests on the prototype are performed to derive the force-displacement relationship. The averaged curve and corresponding transition sequences are presented in Fig. 3*B* and *D*.

As a comparison, simulated force-displacement curves as well as transition sequences are derived in Fig. 3C and *E*. The corresponding parameters adopted in the simulation are $\mu=20$, $\theta_{A\alpha}^0 = -\pi/3$, $k_1^* = 80k$ and $k_2^* = 100k$, with stress-free configuration '1-b-0'.

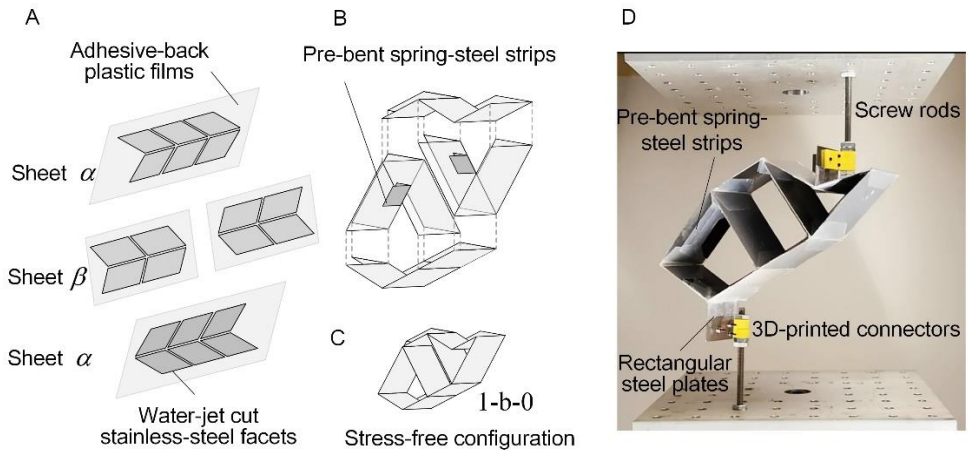

**Fig. S5** prototype of the origami cell. (A-C) illustration of the prototyping method. (D) Experimental setup.

## S6. Dynamic modeling

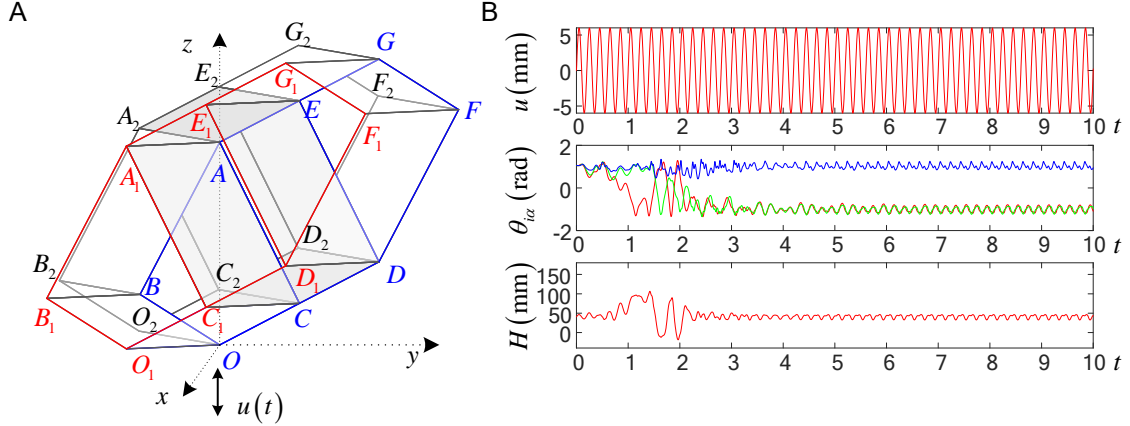

**Fig. S6** Dynamic modeling of the SMOV structure. (A)Vertices of a single SMOV cell. (B) time histories of the excitation and the response

The structure is made up of homogeneous facets with material density $\rho$. Based on the kinematics, the coordinates of the vertexes on the middle layer (colored by blue in Fig. S6*A*) are easily obtained in Table S1.

**Table S1**. Coordinates of the vertexes on the middle layer

| Vertexes $i$ | $O$ | $A$ | $B$ | $C$ | $D$ | $E$ | $F$ | $G$ |
|---|---|---|---|---|---|---|---|---|
| $X_i$ | 0 | 0 | 0 | 0 | 0 | 0 | 0 | 0 |
| $Y_i$ | 0 | 0 | $-\frac{1}{2}L_A$ | $\frac{1}{2}L_A$ | $\frac{1}{2}(L_A+L_B)$ | $\frac{1}{2}L_B$ | $\frac{1}{2}L_B+L_C$ | $\frac{1}{2}(L_B+L_C)$ |
| $Z_i - u(t)$ | 0 | $H_A$ | $H_{A\alpha}$ | $H_{A\alpha}$ | $H_{A\alpha}+H_{B\alpha}$ | $H_A+H_{B\alpha}$ | $H_A+H_{B\alpha}$ | $H_A+H_{B\alpha}+H_{C\alpha}$ |

The coordinates of the vertexes on the top layer, $i_1$ (colored by red in Fig. S6*A*), and the bottom layer $i_2$ (colored by black in Fig. S6*A*) ($i=O,A,B,\cdots,G$) are

$$X_{i_1} = X_i + \frac{1}{2}W_A,\ Y_{i_1} = Y_i - \sqrt{b_\alpha^{\,2} - \left(\frac{1}{2}W_A\right)^2},\ Z_{i_1} = Z_i,$$
$$X_{i_2} = X_i - \frac{1}{2}W_A,\ Y_{i_2} = Y_i - \sqrt{b_\alpha^{\,2} - \left(\frac{1}{2}W_A\right)^2},\ Z_{i_2} = Z_i. \quad [12]$$

The total kinetic energy of the structure is the summation of the kinetic energy of each facet, which can be calculated by integrating the kinetic energy in the X, Y, and Z direction, respectively. Owing to the geometric symmetry, we take half of the structure for analysis, and their kinetic energies can be expressed as

$$T_{OO_1B_1B} = \frac{1}{2}\rho\sin\gamma_\alpha \int_0^{a_\alpha}\int_0^{b_\alpha} \left(\frac{d}{dt}\left(\overrightarrow{OO} + \frac{l_b}{b_\alpha}\overrightarrow{OO_1} + \frac{l_a}{a_\alpha}\overrightarrow{OB}\right)\right)^2 dl_a dl_b,$$

$$T_{OO_1C_1C} = \frac{1}{2}\rho\sin\gamma_\alpha \int_0^{a_\alpha}\int_0^{b_\alpha} \left(\frac{d}{dt}\left(\overrightarrow{OO} + \frac{l_b}{b_\alpha}\overrightarrow{OO_1} + \frac{l_a}{a_\alpha}\overrightarrow{OC}\right)\right)^2 dl_a dl_b,$$

$$T_{AA_1B_1B} = \frac{1}{2}\rho\sin\gamma_\beta \int_0^{a_\beta}\int_0^{b_\beta} \left(\frac{d}{dt}\left(\overrightarrow{OA} + \frac{l_b}{b_\beta}\overrightarrow{AA_1} + \frac{l_a}{a_\beta}\overrightarrow{AB}\right)\right)^2 dl_a dl_b,$$

$$T_{AA_1C_1C} = \frac{1}{2}\rho\sin\gamma_\beta \int_0^{a_\beta}\int_0^{b_\beta} \left(\frac{d}{dt}\left(\overrightarrow{OA} + \frac{l_b}{b_\beta}\overrightarrow{AA_1} + \frac{l_a}{a_\beta}\overrightarrow{AC}\right)\right)^2 dl_a dl_b,$$

$$T_{CC_1D_1D} = \frac{1}{2}\rho\sin\gamma_\alpha \int_0^{a_\alpha}\int_0^{b_\alpha} \left(\frac{d}{dt}\left(\overrightarrow{OC} + \frac{l_b}{b_\alpha}\overrightarrow{CC_1} + \frac{l_a}{a_\alpha}\overrightarrow{CD}\right)\right)^2 dl_a dl_b, \quad [13]$$

$$T_{AA_1E_1E} = \frac{1}{2}\rho\sin\gamma_\alpha \int_0^{a_\alpha}\int_0^{b_\alpha} \left(\frac{d}{dt}\left(\overrightarrow{OA} + \frac{l_b}{b_\alpha}\overrightarrow{AA_1} + \frac{l_a}{a_\alpha}\overrightarrow{AE}\right)\right)^2 dl_a dl_b,$$

$$T_{DD_1E_1E} = \frac{1}{2}\rho\sin\gamma_\beta \int_0^{a_\beta}\int_0^{b_\beta} \left(\frac{d}{dt}\left(\overrightarrow{OD} + \frac{l_b}{b_\beta}\overrightarrow{DD_1} + \frac{l_a}{a_\beta}\overrightarrow{DE}\right)\right)^2 dl_a dl_b,$$

$$T_{DD_1F_1F} = \frac{1}{2}\rho\sin\gamma_\beta \int_0^{a_\beta}\int_0^{b_\beta} \left(\frac{d}{dt}\left(\overrightarrow{OD} + \frac{l_b}{b_\beta}\overrightarrow{DD_1} + \frac{l_a}{a_\beta}\overrightarrow{DF}\right)\right)^2 dl_a dl_b,$$

$$T_{GG_1E_1E} = \frac{1}{2}\rho\sin\gamma_\alpha \int_0^{a_\alpha}\int_0^{b_\alpha} \left(\frac{d}{dt}\left(\overrightarrow{OG} + \frac{l_b}{b_\alpha}\overrightarrow{GG_1} + \frac{l_a}{a_\alpha}\overrightarrow{GE}\right)\right)^2 dl_a dl_b,$$

$$T_{GG_1F_1G} = \frac{1}{2}\rho\sin\gamma_\alpha \int_0^{a_\alpha}\int_0^{b_\alpha} \left(\frac{d}{dt}\left(\overrightarrow{OG} + \frac{l_b}{b_\alpha}\overrightarrow{GG_1} + \frac{l_a}{a_\alpha}\overrightarrow{GCF}\right)\right)^2 dl_a dl_b,$$

Therefore, the total kinetic energy of the structure yields

$$T = 2\left(T_{OO_1B_1B} + T_{OO_1C_1C} + T_{AA_1B_1B} + T_{AA_1C_1C} + T_{CC_1D_1D} + T_{AA_1E_1E} + T_{DD_1E_1E} + T_{DD_1F_1F} + T_{GG_1E_1E} + T_{GG_1F_1G}\right) \quad [14]$$

The governing equation of motion for the origami structure can be constructed by the Lagrange equation for the general case

$$\frac{d}{dt}\left(\frac{\partial(T-\Pi)}{\partial\dot{\theta}_{i\alpha}}\right) - \frac{\partial(T-\Pi)}{\partial\theta_{i\alpha}} = Q_i, \quad [15]$$

where $Q_i$ denotes the nonconservative generalized force associated with the generalized coordinate $\theta_{i\alpha}$ ($i = A, B, C$). Considering the viscous damping, it can be written as

$$Q_i = c\dot{\theta}_{i\alpha}, \quad [16]$$

Substituting the kinetic energy, potential energy, and the generalized forces of the SMOV cell into the Lagrange equation and performing necessary simplification, the equation of motion of the SMOV cell can be obtained.

## S7. Dynamic transforming

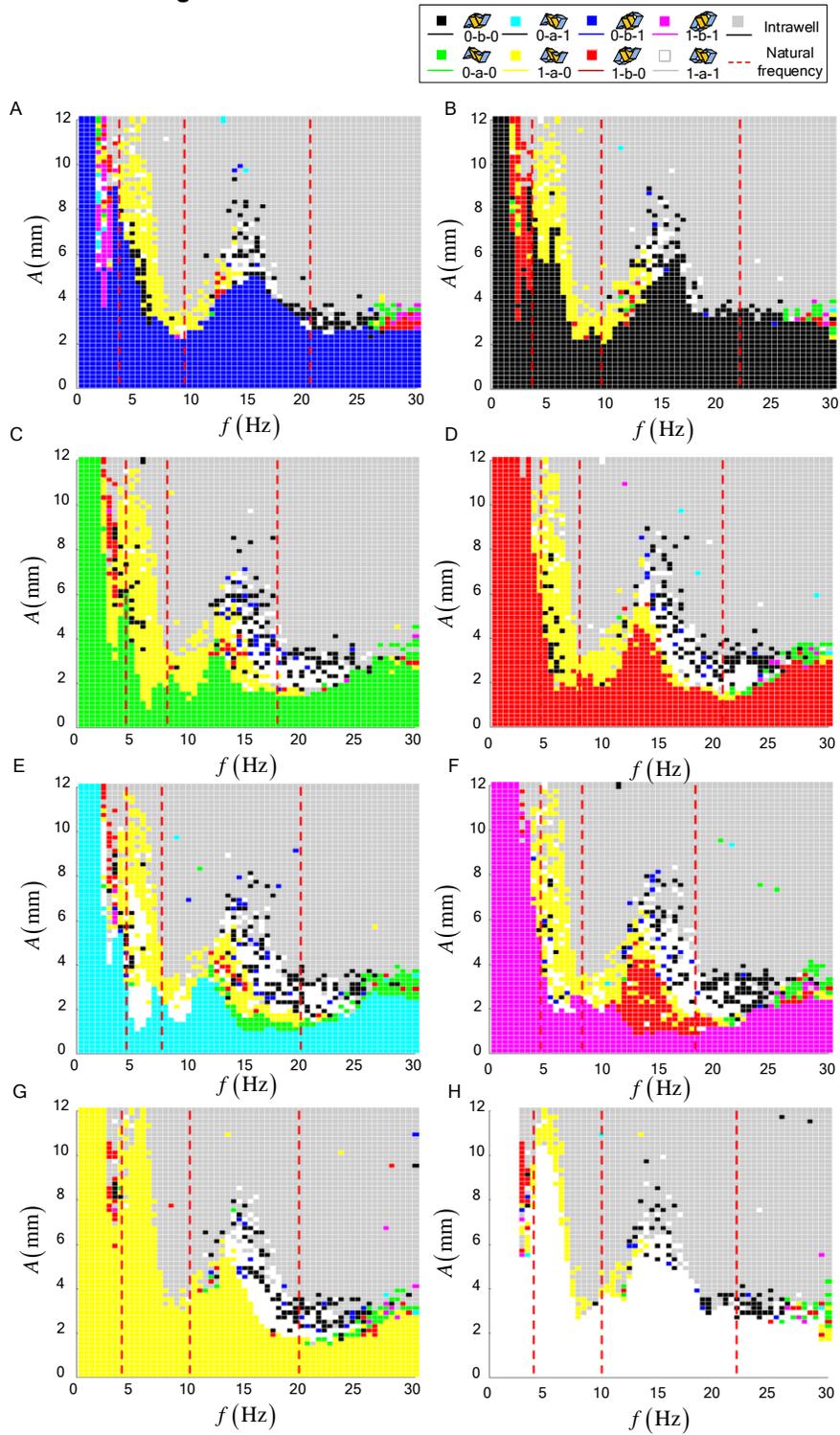

**Fig. S7** Transition sequences with periodic dynamic control. Correlations between the dynamic transitions and the external excitations, i.e., excitation frequency and amplitude. The initial state is at (A) '0-b-1', (B) '0-b-0', (C) '0-a-0', (D) '1-b-0', (E) '0-a-1', (F) '1-b-1', (G) '1-a-0', (H) '1-a-1' configuration with zero initial velocity, and the dashed lines are natural frequencies of the linearized model.

In the simulation, we set $\rho = 7.85\ g/cm^3$, which is the density of steel. The damping coefficient is adopted as $c = 50\ kg \cdot m/s$. Other parameters keep the same as that in the quasi-static experiment. The initial state is set at one of the stable configurations with zero velocity. An example is illustrated in Fig. S6*B*, a sinusoidal excitation with an amplitude of 6mm and frequency of 5Hz is applied. The time history response regarding the three folding angles is presented in the second row. At time instant 0, all the three folding angles are about $\pi/3$, which corresponds to the '0-b-0' stable configuration. However, folding angles of units A and B will become positive in the steady-state, implying that the structure is transformed to oscillate around the configuration '1-a-0'.

Besides, we linearized the dynamic model around its 8 stable configurations. The corresponding natural frequencies are presented in Table S2, which are also plotted with dashed red lines on each map in Fig. S7.

**Table. S2** Natural frequencies at each stable configuration

|  | 0-b-0 | 0-b-1 | 0-a-0 | 1-b-0 | 0-a-1 | 1-b-1 | 1-a-0 | 1-a-1 |
|---|---|---|---|---|---|---|---|---|
| **1st (Hz)** | 3.67 | 3.77 | 4.41 | 4.53 | 4.47 | 4.57 | 4.04 | 3.97 |
| **2nd (Hz)** | 9.75 | 9.52 | 8.06 | 7.95 | 7.58 | 8.28 | 10.06 | 9.99 |
| **3rd (Hz)** | 21.88 | 20.59 | 17.79 | 20.60 | 19.86 | 18.22 | 19.71 | 21.91 |

## S8. Mechano-logic with the transition sequences

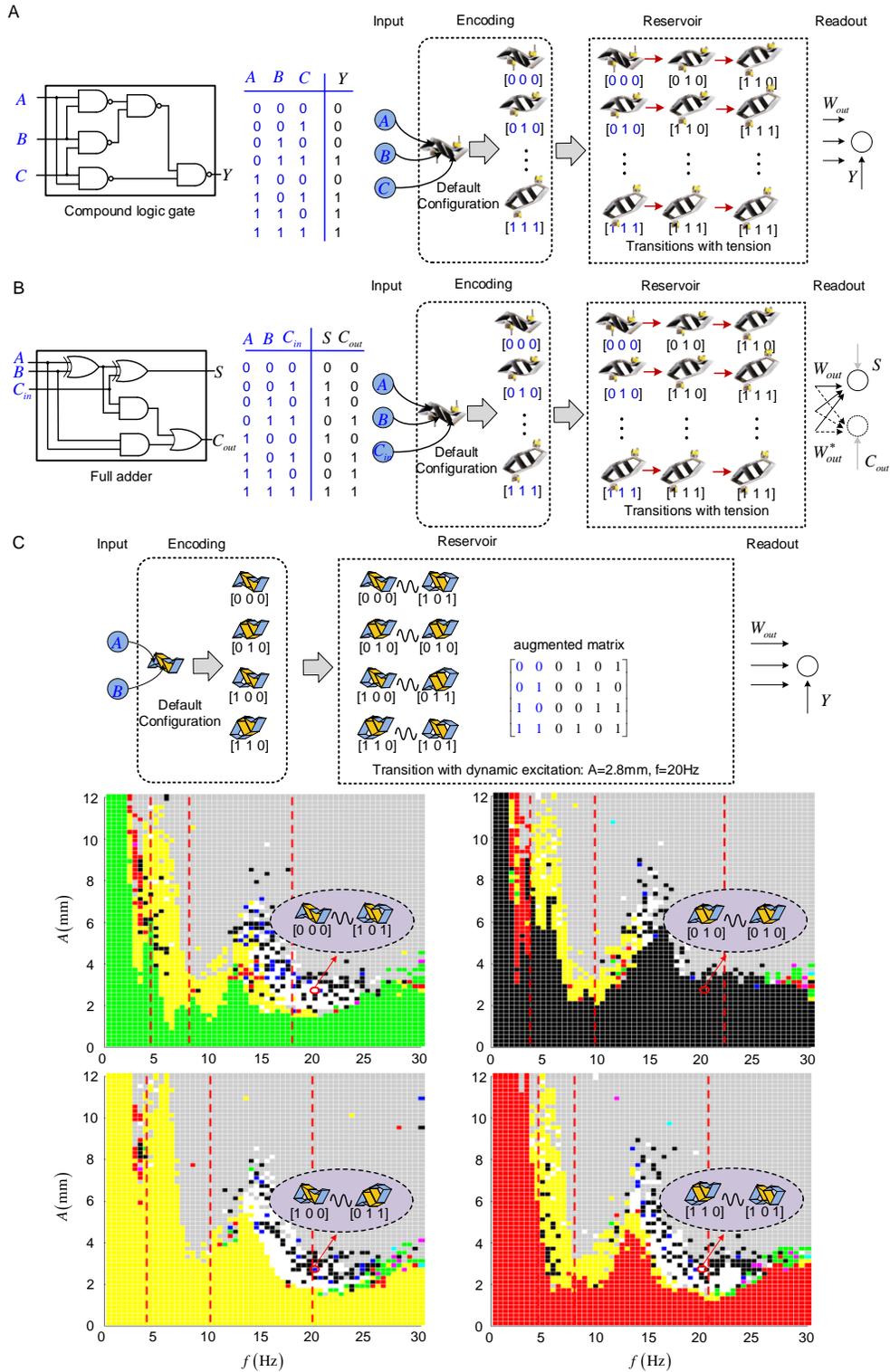

**Fig. S8** The architecture of developing (A) compound logic gate and (B) a full adder with the SMOV cell. (C)The architecture of using the dynamic transition sequences (i.e., the bottom four figures with fixed excitation frequency of 20Hz and amplitude of 2.8mm) for mechano-logic.

**Movie S1. Guiding transition front**

Experimental demonstration of the quasi-static transition sequences. The prototype of the SMOV cell is connected to the Instron machine with a 3D-printed connector. Six complete extension and compression tests starting from different stable configurations are performed. The force-displacement curves, as well as the transitions, are presented during the loading, while the complete transition map is given at the end of the movie.